\documentclass[12pt,a4paper]{article}
\usepackage[dvips]{graphicx}
\newcommand{\bea}{\begin{equation}}
\newcommand{\eea}{\end{equation}}
\newcommand{\ber}{\begin{eqnarray}}
\newcommand{\eer}{\end{eqnarray}}

\newcommand{\p}{\partial}

\newcommand{\f}{\frac}

\begin{document}
\title{1D spirals: is multi stability essential?}
\author{A. Bhattacharyay\footnote {Email: arijit@pd.infn.it} \\ Dipartimento di Fisika 'G. Galilei'\\ Universit\'a di Padova\\ Via Marzolo 8, 35131 Padova\\Italy\\}

\date{\today}
\maketitle
\begin{abstract}
The origin of 1D spirals or antisymmetric 1D pulses is explaind so far on the basis of multistability of spatially inhomogeneous and temporally oscillatory phases and so called nonvariational effects. Thus, coupled amplitude equations which are valid near a co-dimention 2 point and provides with the necessary multistable environment are commonly in use in the numerical calculations to generate such structures. In the present work we analytically show that  
a complex Ginzburg-Landau type amplitude equation which is valid in the Hopf region of phase space near an instability threshold admits solutions like antisymmetric pulses traveling in alternate directions from a core. The pulses can have well defined spatial profile like Hermite polynomial of order unity on the side of the direction of its motion. Thus, a pulse moving in the positive direction of x-axis from near a core at origin will be a peak while that moving in the negative direction is a trough and a temporal oscillation of such structures would give the impression of alternative peak and troughs are moving appart being generated from a common core..\\  
PACS number(s): 87.10.+e, 47.70.Fw
\end{abstract}
\par
 One dimensional spiral pattern, experimentally observed in CIMA reaction is a fascinating phenomenon of nonlinear waves. As described in the experimental part of \cite{per} that at first a transition between a stationary periodic and a traveling wave like Hopf mode is observed by lowering the starch concentration in the chemical reactor. Now, keeping a low starch concentration and by increasing the malonic acid concentration a similar transition takes place which makes an almost 1D Turing state loose its stability to a Hopf state when malonic acid concentration is doubled. It has been observed in this experiment that when the Hopf state takes up from the Turing state very often there remain a few spots, reminiscent of the previous Turing state, acting as the source of one dimensional anti synchronous wave trains. Bands of maximum intensity spreads alternatively toward right and left directions with a time delay. As experiment suggests, this phenomenon is essentially one dimensional and variation of two parameters like concentrations of starch and malonic acid indicates proximity to a co-dimension 2 point. On the basis of this observation a theory has been developed being based on multisability and non variational effects. Such localized structures near a Hopf-Turing instability boundary have been analyzed in many subsequent papers [2-9]. A common belief is that in the range of parameters far away from depinning transitions a pinned Turing state in a global traveling wave phase helps sustain 1D spirals by accepting one on one side of it while leaving the other on the other side. In such a context it is interesting to see if there are other mechanism that can explain the generation of antisymmetric 1D waves from a core without the consideration of multi stability of states. 
\par
In the present paper we are going to show an alternative scenario where the multi stability is not an essential ingredient to make asymmetric outgoing pulses generate from a source. Here, such 1D asymmetric moving pulses are shown as a steady state solution on uniformly moving frame to a complex Ginzburg-Landau type amplitude equation. In a purely Hopf environment where such an equation is applicable and there is no coupling to typical Turing type amplitude equation, small sources can exist which separates two regions where alternative peak and trough of chemical concentrations can move in outward directions from the source. In the present work we emphasize on the fact that generation of 1D spirals can definitely be understood on the basis of a single amplitude equation analysis which is generally ignored with a priori consideration of such things as non variational. On a moving frame of reference the complex Ginzburg-Landau type amplitude equation, applicable to pure Hopf regime, admits steady local structures which have the spatial profile like quantum mechanical harmonic oscillator steady states (Hermite polynomials). To get such linear solutions the nonlinearity is considered negligible which in tern allows only two orders of the polynomial namely zeroth and the first. These two orders are quite different in symmetry and helps explain the antisymmetric effect. These local structures can oscillate when formed in a Hopf environment. It would be shown that on the moving frame the origins of such localized structures are a little shifted on a direction toward which it moves where the amount of shift is proportional to the velocity of the frame. Thus, by considering two moving frames one toward right and the other toward left of some central core one can easily think about a pair of such localized structures moving opposite to each other starting from two points a little right and a little left of the actual origin on lab frame. Moreover, it would be shown that such structures are well defined on the upper(lower) side of their center of a right(left) words moving frame. The inner sides of both of them being less well defined we would better observe the positive x part of the rightwards one and the negative x part of the left moving one and this can give the impression of an antisymmetric wave being generated from a core. thus, when a concentration peak is moving toward right a trough moves toward left with the same velocity.

\par
Let us take the amplitude equation in the form 
\ber\nonumber
\f{\p H}{\p t} &=& (\mu_r+i\mu_i)H + (D_r+iD_i)\f{{\p}^2 H}{{\p x}^2} -(\beta_r+i\beta_i){\vert{H}\vert}^2 H \\\nonumber
\eer
where $H$ is the slow and large scale dependent amplitude of the Hopf mode.

Now, Taking a Hopf phase of the form $H=He^{i\omega t}$ where $H$ is a function of space and time we get,
\ber\nonumber
i\omega H+\f{\p H}{\p t} &=& (\mu_r+i\mu_i)H + (D_r+iD_i)\f{{\p}^2 H}{{\p x}^2} -(\beta_r+i\beta_i){\vert{H}\vert}^2 H \\\nonumber
\eer
After shifting to moving frame with the transformation $y=x-vt$ where $v$ is the velocity of the frame we readily arrive at
\ber\nonumber
(\mu_r+i\mu_i-i\omega)H + (D_r+iD_i)\f{{\p}^2 H}{{\p y}^2} +v\f{\p H}{\p y}-(\beta_r+i\beta_i){\vert{H}\vert}^2 H =0\\\nonumber
\eer
Let us now assume that $H$ has a local form like $He^{\f{-y^2}{2b}}$. With this consideration the above equation changes to
\ber\nonumber
&&(D_r+D_i)\left[\f{{\p}^2 H}{{\p y}^2}e^{\f{-y^2}{2b}}-\f{2y}{b}\f{\p H}{\p y}e^{\f{-y^2}{2b}}+\f{y^2H}{b^2}e^{\f{-y^2}{2b}}-\f{H}{b}e^{\f{-y^2}{2b}}\right]\\ &&v\left[\f{\p H}{\p y}e^{\f{-y^2}{2b}}-\f{Hy}{b}e^{\f{-y^2}{2b}}\right]+(\mu_r+i\mu_i-i\omega)He^{\f{-y^2}{2b}} +(\beta_r+i\beta_i)H^3e^{\f{-3y^2}{2b}} =0\\\nonumber
\eer
Since we are interested in a form of $H$ which is equal to $ye^{\frac{-y^2}{2b}}$ we can safely drop the nonlinear term in $H$. This approximation will be automatically justified latter. A point to be noted here is that, the linear solution is everywhere valid within the envelop $e^{\frac{-y^2}{2b}}$ Since, near origin $y$ is small $H$ is small and if the envelop is not too wide, away from the origin $H$ is also small and the nonlinear term do not count.
\par
The Equations coming out by equating real and imaginary parts will now look like
\ber\nonumber
D_r\left[\f{{\p}^2 H}{{\p y}^2}-\f{2y}{b}\f{\p H}{\p y}+\f{y^2H}{b^2}-\f{H}{b}\right]+v\left[\f{\p H}{\p y}-\f{Hy}{b}\right]+\mu_rH=0\\\nonumber
\eer
and
\ber\nonumber
D_i\left[\f{{\p}^2 H}{{\p y}^2}-\f{2y}{b}\f{\p H}{\p y}+\f{y^2H}{b^2}-\f{H}{b}\right]+v\left[\f{\p H}{\p y}-\f{Hy}{b}\right]+(\mu_i-\omega)H=0\\\nonumber
\eer
They look almost the same except for the last terms. Now, considering one that comes from the real parts and rearranging the terms we get
\ber\nonumber
D_r\f{{\p}^2 H}{{\p y}^2}-\f{2D_r}{b}(y-\f{vb}{2D_r})\f{\p H}{\p y}+\f{D_r}{b^2}(y-\f{vb}{2D_r})^2H-\f{b^2v^2}{4D_r}H+(\mu_r-\f{D_r}{b})H=0\\\nonumber\\
\eer
The above form of the equation suggests a natural shift of the origin by an amount $\f{vb}{2D_r}$. The shift in the origin will be more with faster moving frames and with a wider spread of the localized solution. Let us effect this shift by going to $z=y-\f{vb}{2D_r}$ and after effecting that to look at around this new origin one can easily drop the term containing $\f{z^2}{b^2}$ with a promise that we will consider only linear terms in the solution of this equation. So, the region of interest has a spread of $b$ on the far side of this shifted origin because of the fact that the nonlinear term in $H$ is more relevant towards the inner side. Now, the equation takes the form
\ber\nonumber
D_r\f{{\p}^2 H}{{\p z}^2}-\f{2D_rz}{b}\f{\p H}{\p z}+(\mu_r-\f{D_r}{b}-\f{b^2v^2}{4D_r})H=0\\\nonumber
\eer
\par
Now, the last transformation required is going to a larger scale defined by $z=\f{z}{\sqrt{b}}$ which will produce the final form of the equation as
\ber\nonumber
\f{{\p}^2 H}{{\p z}^2}-2z\f{\p H}{\p z}+\f{b}{D_r}(\mu_r-\f{D_r}{b}-\f{b^2v^2}{4D_r})H=0\\
\eer
\par
The linear Eq.3 admits solutions which are Hermite polynomials and a solution of order zero is obtained when $(\mu_r-\f{D_r}{b}-\f{b^2v^2}{4D_r})$ is equal to zero. When $\mu_r-\f{D_r}{b}-\f{b^2v^2}{4D_r}=\f{2D_r}{b}$ it admits a solution same as Hermite polynomial of order unity. We would not consider the solutions of higher order than unity since we have ignored the $z^2$ term to arrive at Eq.3 but this will suffice to serve our purpose. Now, consider the steady state solution to Eq.3 in the form of Hermite polynomial of order unity. This profile will be uniformly moving toward right starting from a point separated by $\f{vb}{2D_r}$ right of the stationary origin at time $t=0$. what about the same thing with $v=-v$? There will be a similar local structure moving toward left starting from the left of the stationary origin at a distance $-\f{vb}{2D_r}$ at $t=0$. Thus, we can actually think about separating the space simultaneously into two moving frames on the two sides of the stationary core while on one a pulse moves toward right and on the other it moves toward left. 
\par
Let us think about the effect of the shift in origin by the amount $\f{vb}{2D_r}$ from the center of the envelop given by $e^{\frac{-y^2}{2b}}$ in the case of the right-moving solution. In this case, within the envelop the right hand part of the Hermite polynomial of order unity will be suppressed and the left hand part will be most prominent. So, There will be a trough moving towards right hand side when by the same logic, due to the shift in origin within the left-moving envelop, a peak will be moving towards the laft hand side. It is interesting to see that for a narrow envelop ($b$ small) the suppression of one half of the spatial solution is more prominent and the antisymmetric nature of the solution will be more revealed and at the same time a narrow envelop makes the nonlinear effect even more negligible.  
Now, if we remember the fact that these structures are oscillating with frequency $\omega$, we readily realize that this scenario will be changing with time. So, a peak will follow a trough on the right hand side while a trough will follow a peak on the left after a time $\f{1}{2\omega}$.  
\par
Considering the equation obtained from imaginary part along with that from real part requires simultaneous validity of
\ber\nonumber
(\mu_i-\omega-\f{D_i}{b}-\f{b^2v^2}{4D_i})&=&\f{2D_i}{b}\\\nonumber
(\mu_r-\f{D_r}{b}-\f{b^2v^2}{4D_r})&=&\f{2D_r}{b}\\\nonumber
\eer
So, the solution now can be obtained when 
\ber\nonumber
(\mu_i-\omega)\f{b}{D_i}-\f{b^3v^2}{4D_i^2}=\mu_r\f{b}{D_r}-\f{b^3v^2}{4D_r^2}
\eer
meaning that, on the straight line in $\mu_r-\mu_i$ plane for given other parameters, $\omega$ being the frequency of oscillation of moving local structures can adjust itself to lock to the solution of order unity.
\par
In the situation when $D_r$ and $D_i$ are negative, one has to add positive quartic terms to the amplitude equation to make the system saturate. Thus, in such a situation the system is sub critical. Since now $\f{vb}{2D_r}$ changes sign the right moving pulse will have its origin on the left of the zero of $y$ and vice versa. In this situation the assumption of two independent frames on either sides breaks down and as a result such phenomenon are strictly supercritical. We would like to mention that super criticality has also been taken as a condition in numerical simulations to show generation of 1D spirals on the basis of multi stability \cite{per}. As the present theory shows, one can really expect such moving local structures can also be seen in a pure Turing environment near instability boundary where the same form of amplitude equation applies but with real coefficients. The only difference will be that those moving localized structures will be non oscillatory and thus concentration peaks will travel toward right while troughs toward left. Experiments, in this direction, to capture such non oscillatory peaks and troughs moving on the two sides of a core can show the validity of this theory and can verify if the non variational effects are essential or not for such pulse generation.
\par
In the conclusion we would like to say that we have put forward an alternative theory for antisymmetric pulse propagation or 1D spirals which has nothing to do with multi stability of states. The consideration of non variational effect comes out to be nonessential in view of the fact that such pulses can arise in a pure Turing type amplitude equation which has a potential form. There are certainly inherent mechanism other than multi stability of phases which is also capable of showing the phenomenon of 1D spiral.

\end{document}